\documentclass[twoside,twocolumn]{article}

\usepackage{blindtext} 

\usepackage[sc]{mathpazo} 
\usepackage[T1]{fontenc} 
\usepackage{microtype} 

\usepackage[english]{babel} 

\usepackage[hmarginratio=1:1,top=32mm,columnsep=20pt]{geometry} 
\usepackage[hang, small,labelfont=bf,up,textfont=it,up]{caption} 
\usepackage{booktabs} 

\usepackage{lettrine} 

\usepackage{enumitem} 
\setlist[itemize]{noitemsep} 

\usepackage{abstract} 

\usepackage{titlesec} 
\renewcommand\thesection{\Roman{section}} 
\renewcommand\thesubsection{\roman{subsection}} 
\titleformat{\section}[block]{\large\scshape\centering}{\thesection.}{1em}{} 
\titleformat{\subsection}[block]{\large}{\thesubsection.}{1em}{} 

\usepackage{fancyhdr} 
\usepackage{titling} 

\usepackage{hyperref} 

\usepackage{natbib,twoopt}
\usepackage{subfigure}
\usepackage{cases}
\usepackage{graphicx}
\usepackage{txfonts}

\pretitle{\begin{center}\Huge\bfseries} 
\posttitle{\end{center}} 
\title{Statistics of Photospheric Supergranular Cells Observed by SDO/HMI} 
\author
{Majedeh Noori$^{1}$, Mohsen Javaherian$^{2}$, Hossein Safari$^{1\ast}$, Hamid Nadjari$^{1}$\\
\\
\normalsize{$^{1}$Department of Physics, University of Zanjan,}\\
\normalsize{University Blvd., Zanjan, IRAN, P. O. Box:  45371-38791.}\\
\normalsize{$^{2}$Research Institute for Astronomy and Astrophysics of Maragha (RIAAM),}\\
\normalsize{Maragha, IRAN, P. O. Box: 55134-441.}\\
\\
\normalsize{$^\ast$Hossein Safari; E-mail: safari@znu.ac.ir}
}
\date{\today} 
\renewcommand{\maketitlehookd}{%
\begin{abstract}
{\textit{\textsf{Aims:}} The statistics of the photospheric granulation pattern are investigated using} continuum images observed by Solar Dynamic Observatory (SDO)/Helioseismic and Magnetic Imager (HMI) taken at 6713~\AA.

{\textit{\textsf{Methods:}} The supergranular boundaries can be extracted by tracking photospheric velocity plasma flows. The local ball-tracking method is employed to apply on the HMI data gathered over the years 2011-2015 to estimate the boundaries of the cells. The edge sharpening techniques are exerted on the output of ball-tracking to precisely identify the cells borders. To study the fractal dimensionality (FD) of supergranulation, the box counting method is used.}

{\textit{\textsf{Results:}} We found that both the size and eccentricity follow the log-normal distributions with peak values about 330 Mm$^2$ and 0.85, respectively. The five-year mean value of the cells number appeared in half-hour sequences is obtained to be about 60 $\pm$ 6 within an area of $350^{\prime\prime}\times350^{\prime\prime}$. The cells orientation distribution presents the power-law behavior.}

{\textit{\textsf{Conclusions:}} The orientation of supergranular cells ($O$) and their size ($S$) follows \textbf{a} power-law function as $|O| \propto S^{9.5}$. We found that the non-roundish cells with smaller and larger sizes than 600 Mm$^2$ are aligned and perpendicular with the solar rotational velocity on the photosphere, respectively. The FD analysis shows that the supergranular cells form the self-similar patterns.}

\end{abstract}

\textit{\textsf{Sun: photosphere - Sun: granulation - Methods: data analysis - Techniques: image processing}}
}
\begin{document}
\maketitle


\section{Introduction}

\large{The solar granulation is the upper side of convective cells produced based on traveling hot plasma currents from the solar interior (convective zone) to the photosphere \citep{Priest2014}. Hot plasma rises up to the surface and transfers energy. Then, cold plasma returns to the interior within the dark boundaries. The granulation is a turbulent process done by merging new grains and splitting old ones \citep{Javaherian2014}. The size of supergranules covers the extended range of scales, with a typical diameters of 20-70 Mm \citep{Priest2014, Margarita}. The horizontal velocity of the plasma flows from the cell centers toward the edges are estimated to be $0.2-0.5$ km s$^{-1}$ \citep{Simon1968, Margarita}.

Granulation process is linked to the magnetic flux distributed ubiquitously throughout the solar surface. It has been shown that the emergence of magnetic flux is related to the cells where the plasma flows positively diverge \citep{Stangalini2013}. The large-scale type of granulation occurs in supergranular cells which is one of the characteristics of the quiet Sun \citep{Priest2014}. Studying the statistics of the surface magneto-convective features lead to better understanding of their evolution.

Photospheric convective pattern highlights the evolution of other phenomena, such as coronal bright points, magnetic cancelation, nanoflares and network flares in different layers \citep{Ryutova, Taj, Yousefzadeh2016}. \citet{Tian2010} used data in various passbands to find out a correlation between the horizontal velocities of plasma in both photospheric and chromospheric supergranulation. In one of the statistical works, the average diameter and lifetime of supergranular cells were found to be 25 Mm and 1.5 days, respectively \citep{Roudier2014}. The relation between the cell size and the magnetic field is unclear; but, in recent works done based on local correlation tracking (LCT), the cell size and velocity are linked to the intensity map of both supergranular vertical and horizontal flows \citep{Rincon2017}. This indirect relation between the cell size and magnetic flux inside the cells leads to anti-correlated dependence, so the large cells can be emerged where the local magnetic field is weak.

The relation between supergranular attributes and the solar cycle, specially, the cell size and intensity variation have been studied by \citet[]{Meunier2008}. They used Michelson Doppler Imager \citep[MDI:][]{Scherrer} onboard on Solar and Heliospheric Observatory (SOHO), to show that the size of supergranules are smaller at the maximum of solar activity. The intensity variation of supergranular cells from center to boundary is comparable with that of obtained for granules \citep{Del Moro2007}.

\citet[]{Meunier2007} have presented results about cell-size distribution and found a correlation between horizontal velocity of plasma and supergranular radius extracted from MDI/SOHO. In recent decades, by increasing received data taken from ground-based and space telescopes, the automatic detection methods are intensively expanded to analyze data and extract statistics with higher accuracy \citep[e.g., see][]{Aschwanden2010, Alipour2015, Arish2016, Javaherian2017}, and also, reducing costs of data classification. So, the numerous methods are progressed to extract supergranular cells from intensity continuum images. One of the important approaches, known as \textit{local correlation tracking} (LCT), is employed to recognize the boundaries based on flux current in data \citep{Papadimitriou2006}. The velocity of pixels of interests are computed by capturing the transform motion of pixels in two consecutive frames. A two-dimensional (2-D) flow field can represent the boundaries of cells as the output of the method. \citet{Fisher2008} extended the LCT method with application of the fourier transform named \textit{fourier local correlation tracking} (FLCT).

One of the promising algorithms for determining the supergranular boundaries is \textit{ball-tracking method}. This method considers imaginary balls on the gridded surface independently moving with plasma flows \citep{Potts2004}. According to the intensities, the balls move in directions to settle in local minima where the boundaries are elongated. Using balls tracking, the locations of some coronal small-scale features (bright points, mini-coronal mass ejections, etc.) were carried out by \citet[]{Innes}, \citet[]{Yousefzadeh2016}, and \citet[]{Honarbakhsh}.

We investigate the supergranules morphological parameters and velocities during a five-year period of the solar activity. So, the statistical parameters of supergranular cells, such as size$-$frequency distribution, fractal dimension (FD), orientation, and their eccentricities are studied. Moreover, the correlations between quantified parameters and the solar activity are computed. We used data recorded by Solar Dynamic Observatory (SDO)$/$ Helioseismic and Magnetic Imager (HMI) taken at 6173~\AA.

The layout of this article is as follows: the description of data sets is given in Sect. \ref{data}. The brief review of the methods are prepared in Sect. \ref{Method}. The results are presented in Sect. \ref{Results}. Concluding remarks are explained in Sect. \ref{Con}.
}
\section{Description of Datasets}\label{data}

Solar Dynamic Observatory (SDO) utilized one of the three instruments named Helioseismic and Magnetic Imager \citep[HMI:][]{Schou2012} to investigate the photospheric oscillations and magnetic fields \citep{Wachter2012}. So, to study the supergranular cells using photospheric continuum images, we employed the high-spatial and temporal resolution data recorded by HMI. The HMI provides different level of full-disk images in the Fe~{\small{I}} absorption line at 6173~\AA~ with a resolution of $0.50\pm0.01$ arcsec and cadence of 45 seconds. Some corrections, such as exposure time, dark current, flat field, and cosmic-ray hits, are done in level-1 data.

\begin{figure*}[ht!]
    \centering
    \begin{subfigure}[b]
        \centering
        \includegraphics[width=8.2cm]{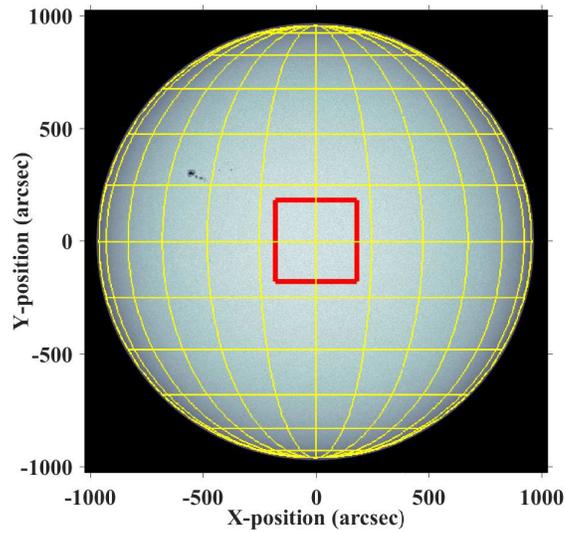}
    \end{subfigure}%
    \begin{subfigure}[b]
        \centering
        \includegraphics[width=8.4cm]{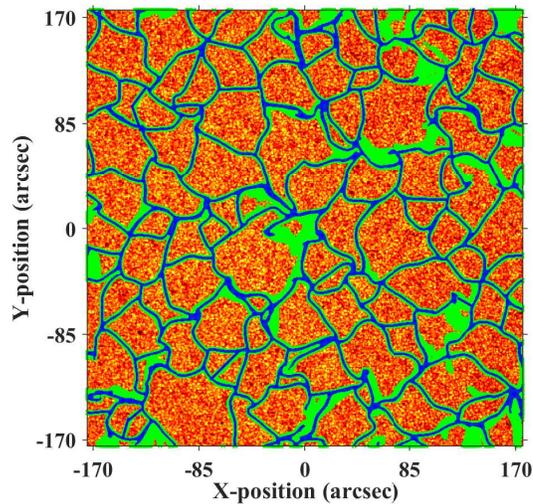}
    \end{subfigure}

    \caption{The continuum SDO/HMI full-disk image of the Sun recorded on 30 December 2015 (00:00-00:30 UT) taken at 6173~\AA~(left panel). The red rectangle is selected with area of $350^{\prime\prime}\times350^{\prime\prime}$ (region centered with longitudes $\pm 11^\circ$ around the central meridian, and the latitudes limited in $\pm 11^\circ$ around the equator). The output of the ball-tracking method is shown with green face (right panel). The blue edges are the cells border obtained by the morphological filters (see text).}\label{fig1}
\end{figure*}

For our purpose, we used $30$-minute consecutive continuum HMI data with a time lag of 45 seconds in every two days from the year $2011$ to $2015$. Since the measurement of morphological parameters of the supergranules, such as size and orientation, and also, velocity on the surface very sensitive to the projection effect, variation of the solar radial over time, and the $B_0$ angle evolution effect \citep{Roudier2013}, the partial area with a size of $350^{\prime\prime}\times350^{\prime\prime}$ at the solar disk center (a region centered with longitudes $\pm 11^\circ$ around the central meridian, and latitudes limited in $\pm 11^\circ$ around the equator) (Fig. \ref{fig1}, red box) was selected to focus on photospheric flows. To coalign the sequential data to a reference one, all images are derotated using \textsf{drot$_{-}$map.pro} available in the SSW/IDL package.

\section{Methods}\label{Method}

The ball-tracking method, edge sharpening technique, and box-counting algorithm are explained briefly as follows.

{\textbf{\textsf{Ball-tracking$\centerdot$}} One of the applicable methods developed for computing velocity fields is ball-tracking method \citep{Potts2004}. The ball-tracking method is applied on continuum HMI images to track velocity fields. Using fast fourier transform which is a part of the ball-tracking code, photospheric $p$-mode oscillations are attenuated to remove features moving faster than $7$ km~s$^{-1}$. In the code, the constructed data cube with two spatial and one temporal dimension is converted to a cube in feature space of wave vector ($k$) and frequency ($\omega$) by Fourier transform. As photosheric oscillations ($\upsilon_{{\rm cut-off}} = \omega / k$) indicates a cone lateral surface in $k-\omega$ space, we are able to discard components covering motions greater than $\upsilon_{{\rm cut-off}} = 7~km~s^{-1}$ \citep{Title1989, Roudier2013}.

Ball-tracking method delineates velocity fields in a fraction of time by moving spherical balls as float tracers on a surface. In this method, the granulation pattern is considered as a criterion for 3-D tracers on surface. Small floating balls move on the solar photosphere based on bouncy laws. In other words, merging and splitting of granules reveal bumps on surface wherever the time-evolving granulation form ripples on the surface. These movements are very similar to a ball moving on fluid surface. This ball has a given mass and momentum. If granular cells push the ball, the sphere will continue traveling based on incoming force. So, the path and direction of ball are estimated as a function of time \citep{Potts2008}. Both the diameter of balls and average surface penetration depend on the resolution of the image. The radius of ball is chosen to match the typical size of granular cells. The smaller the radius, the cell borders are found with more sensitivity to short wavelength noise. Typical value for the radius is half of center-to-center granular distance. For high resolution HMI data, the ball diameter is around two pixels. Time range of $30$ minutes is selected to prepare images as a datacube. Figure \ref{fig1} (right panel) exhibits the ball-tracking output extracted from HMI data displayed as regions with green face.

{\textbf{\textsf{Edge Detection$\centerdot$}} To sharpen the edges, the binary format of images underwent the bridge algorithm that fill the blank space appeared between unclosed boundaries \citep{Gonzalez2009}. In the next step, to fill remained gaps in boundaries, a decomposition algorithm is used to structuralize edges (if it exists) by image dilation. To speed up performing the binary dilation, erosion procedure is applied on output image and the lanes were nearly closed \citep{Boomgaard1992}. To find perimeter of structures in 2-D images, the algorithm uses connectivities to specify the edges with more width. The output image including boundaries is presented with solid blue lines in Fig. \ref{fig1} (right panel).

{\textbf{\textsf{Box-counting$\centerdot$}} Fractals have geometric repetitive patterns in different scales wherein the whole of structure can be generated by small parts in non-integer dimensions \citep{Aschwanden2011}. One of the methods that estimates the fractal dimension of 2-D data is box-counting \citep{Molteno1993}. This method is applied on image to breaks it into smaller parts with different resolutions step-by-step. In different resolutions, the code considers the boxes consisting of image components. In each step, the size of box ($\epsilon$) and the number of boxes (N($\epsilon$)) is computed. Thus, the image dimension (D) is defined as

\begin{eqnarray}
D = \lim_{\epsilon \to 0}\frac{\log N(\epsilon)}{\log (\epsilon^{-1})}.
\label{e0}
\end{eqnarray}

\section{Results} \label{Results}
To determine the boundaries, an automatic recognition method is applied to HMI intensity-continuum data recorded at 6173~\AA. We focused on rectangular area of $350^{\prime\prime}\times350^{\prime\prime}$ to extract statistical properties of supergranular cells (Fig. \ref{fig1}, left panel). For our purpose, $40$ images (half-hour data) for every two days were gathered with a time lag of $45$ seconds during the years 2011 to 2015.

The log-normal function is fitted to the size-frequency distribution of supergranules, (Fig. \ref{fig2}, upper panel). The log-normally distributed function is given by \citep{Newman2006, Bazar, Aschwanden2015},

\begin{eqnarray}
f(x, \mu, \sigma)=\frac{1}{\sigma x \sqrt{2\pi}}\exp \left( -\frac{(\ln(x)-\mu)^{2}}{2\sigma^{2}}\right),
\label{e1}
\end{eqnarray}
where $\mu$ is the mean value, and $\sigma$ is the standard deviation. The fit parameters $\mu$ and $\sigma$ for overall five-year size-frequency are obtained to be $5.208 \pm 0.038$ and $0.757 \pm 0.031$, respectively, with the peak value of $330$ Mm$^{2}$. The variation of these values is approximately constant during five years (Fig. \ref{fig2}, upper panel).

\begin{figure}[ht!]
\centerline{\includegraphics[height=5cm]{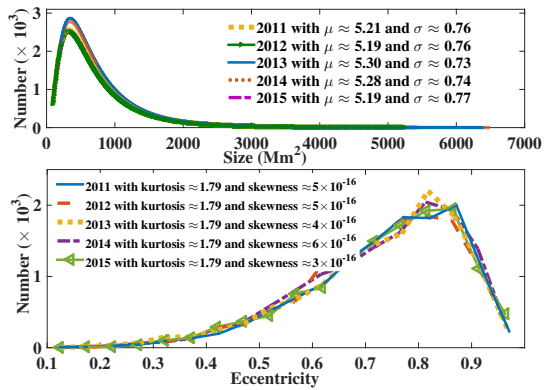}}
\caption{The histograms of cells size (top) and cells eccentricity (bottom) for each year. The log-normal function (Eq. \ref{e1}) is fitted to the size distributions.}\label{fig2}
\end{figure}

Each segmented region can be surrounded by an ellipse that can be attained by morphological moments of a shape \citep{EmreCelebi2005}. Thus, the major and minor axis of contained area is computed to obtain the eccentricity. All cells eccentricities take values ranged from 0 and 1. The value 0 represents the shape as a circle, and 1 returns a segmented shape as line.

The peak value of the eccentricity distribution of supergranular cells is $0.8$. The skewness takes a positive value and kurtosis is 1.8 for overall five years. The skewness and kurtosis of eccentricity distributions are approximately constant during five years (Fig. \ref{fig2}, lower panel).

The cell orientation between the major and horizontal axis (west-east latitude on the solar disk) and an estimation of the error value of the orientation angles are specified for each region (see Appendix A). The orientation distribution of cells follows a power-law fit as $y \propto x^{-\alpha}$, wherein $\alpha$ is the power exponent. Using the methods introduced in \citet{Aschwanden2015} and \citet{SafariFarhad}, the power-law exponents were obtained. In Fig. \ref{fig3}, the fitted power-law function is shown with the exponent \textbf{$\alpha = 0.1613 \pm 0.0416$} which seems to be constant over the period. The errors for the orientations of the cells were in the range of 0.0001$^\circ$ to 1.7321$^\circ$ with mean values of 0.0584$^\circ$.

\begin{figure*}[ht!]
\centerline{\includegraphics[height=7.8cm]{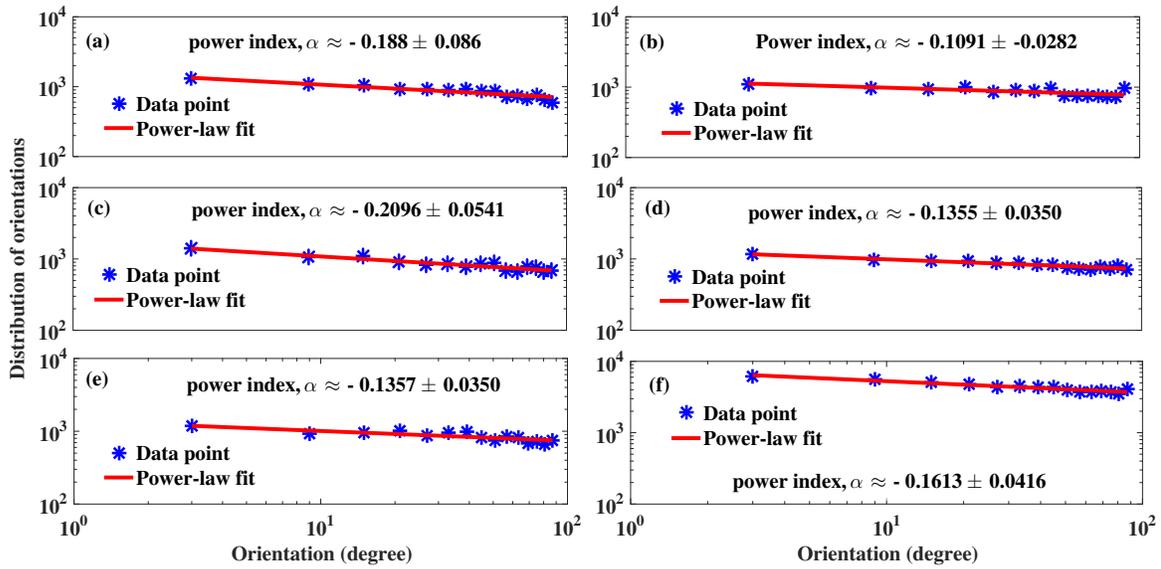}}
    \caption{The orientation distribution of supergranules for the years of $2011-2015$ (a: 2011, b: 2012, c: 2013, d: 2014, e: 2015, f: 2011-2015). The minimum and maximum of orientation errors computed for regions are about 0.0001$^\circ$ and 1.7321$^\circ$, respectively. The maximum error values belong to the cells orientated in $\pm 45^\circ$.}\label{fig3}
\end{figure*}

In Fig. \ref{fig4}, the log-normal function is fitted on the photospheric horizontal velocities of plasma for each year (Fig. \ref{fig4}). The overall five-year parameters $\mu$ and $\sigma$ are equal to $4.634 \pm 0.039$ and $0.556 \pm 0.031$, respectively. The time series of daily (blue) and monthly (red) number of cells for five years are shown in Fig. \ref{fig5}. The mean value is equal to 59.6 $\pm$ 5.62 cells ranged from 43 to 77 in each frame.

\begin{figure}[ht!]
\centerline{\includegraphics[height=6.3cm]{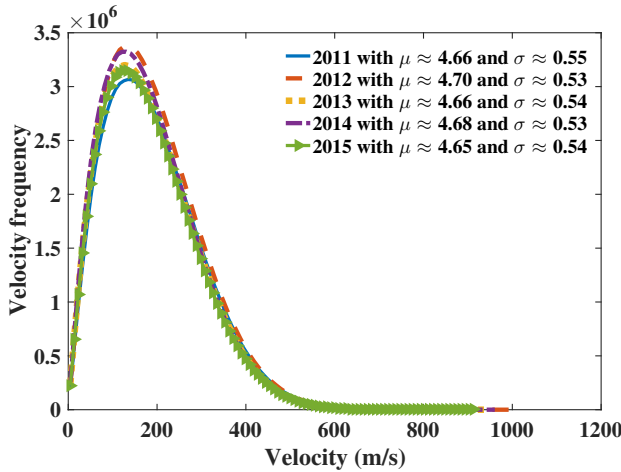}}
\caption{The histograms of horizontal velocities for each year during the years 2011 to 2015.
The log-normal function (Eq. \ref{e1}) is fitted to the distributions.}\label{fig4}
\end{figure}

\begin{figure*}[ht!]
\centerline{\includegraphics[height=6.5cm]{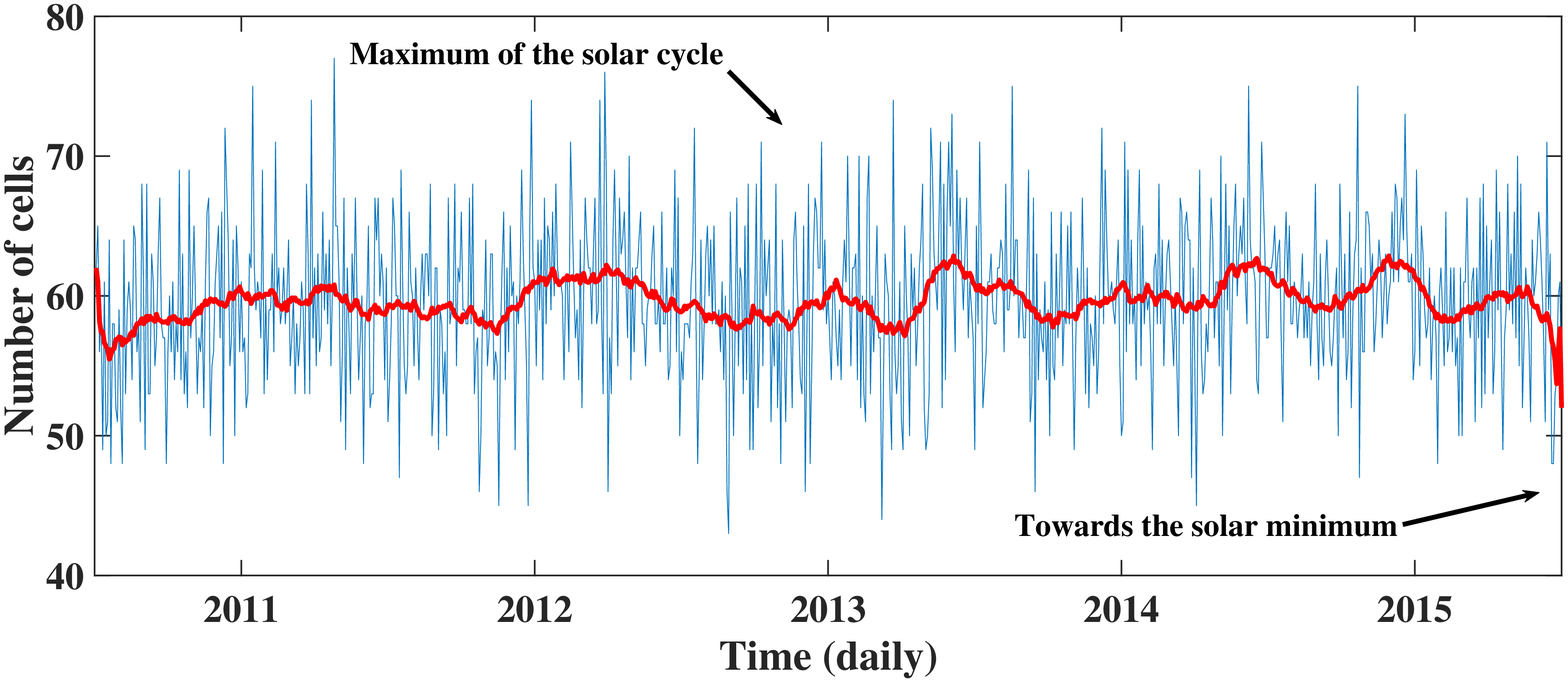}}
\caption{\textbf{Time series of} daily (blue) and monthly smoothed (red) number of cells during the years 2011 to 2015.}\label{fig5}
\end{figure*}

As we see in Fig. \ref{fig6}, the time series of the plasma velocity and the number of sunspots are presented. To test Pearson correlation between time series, we employed the box-Cox transformation to convert the distributions of time series with positive values (number of cells, number of sunspots, etc.) to normal distributions. This transformation considers a range for exponents ($\lambda$) defined in following equation

\begin{subnumcases}{\label{box} Y(\lambda)=}
\frac{y^{\lambda}-1}{\lambda}, ~~~~~ {\rm if} ~~ \lambda \neq 0, \nonumber \\
\log(y), ~~~~~ {\rm if} ~~ \lambda = 0, \nonumber
\end{subnumcases}
where $\lambda$ usually varies from -5 to 5. The optimal value of $\lambda$ is obtained as the best approximation fitted on the normal distribution curve \citep{Everitt2002}. To do this, first, kurtosis for each transformed time series with value about three are chosen for the best $\lambda$. Then, the correlation coefficients are computed for time series \citep{Press2007}.

To validate the values attained by Pearson correlation, a hypothetical test called p-value (probability value) is exploited. It specifies that whether there is a meaningful relation between time series or computed correlation has been occurred by accident. The p-value smaller than 0.05 shows the higher validity of the correlation \citep{Everitt2002}.

As seen in Table \ref{t1}, the correlations between the cells size and orientations, sunspots number and velocities, and also, sunspots number and cells number are about 0.3, 0.2, and 0.3, respectively. There is an anti-correlated behavior between eccentricities and cells size, and also, sunspots number and cells size with about - 0.1 and -0.4, respectively.

\begin{figure*}[ht!]
\centerline{\includegraphics[height=6.5cm]{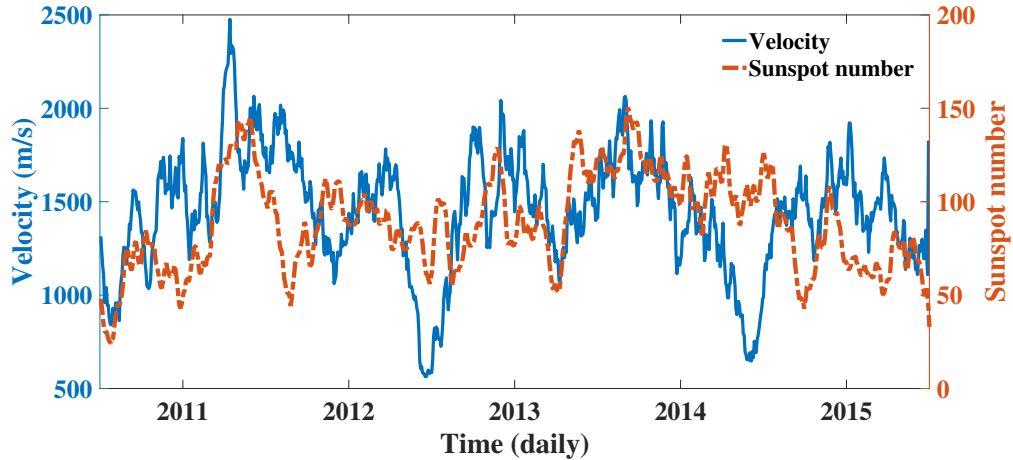}}
\caption{Time series of both smoothed monthly number of sunspots (red) and surface velocities (blue).}\label{fig6}
\end{figure*}

\begin{table*}[ht!]
\begin{center}
\caption{Pearson correlations between statistical parameters.} \label{t1}
\begin{tabular}{l  c  c  c }
\hline
    & $\lambda$ & Correlation & P-value \\ \hline \hline

Orientations and cells size & -0.700 & 0.296 & 0.001 \\ \hline
Eccentricities and cells size & 5.000 & -0.121 & 0.001 \\ \hline
Sunspots number and velocities &  2.400 & 0.244 & 0.001 \\ \hline
Sunspots number and cells size & 0.300 & -0.410 & 0.001 \\ \hline
Sunspots number and cells number & 0.300 & 0.320 & 0.001 \\ \hline
\end{tabular}
\end{center}
\end{table*}

As shown in Fig. \ref{fig7} (upper panel), for eccentricities smaller than 0.2, the sizes are more fluctuated. As we see in Fig. \ref{fig7} (lower panel, blue line), by increasing the size, the orientation rises. The small cells with eccentricities around 0.55 have orientations close to $0^{\circ}$, and with increasing eccentricity, the orientation rises and approaches to 0.8 (Fig. \ref{fig7}, red line in lower panel). The size ($S$) and the orientation ($O$) are related by $|O| \propto S^{9.509 \pm 0.865}$. Relationships between the size of cells, orientation, and eccentricity show that the large cells are commonly included orientations around $45^{\circ}-90^{\circ}$ with shapes similar to ellipse, and the smaller ones are in the range of $0^{\circ}-45^{\circ}$ with mostly non-roundish shapes (Fig. \ref{fig7}). In Fig. \ref{fig8}, the direction of cells (orientations) and surrounded ellipses are displayed by black lines and red ellipses, respectively.

\begin{figure}[ht!]
\centerline{\includegraphics[height=6.3cm]{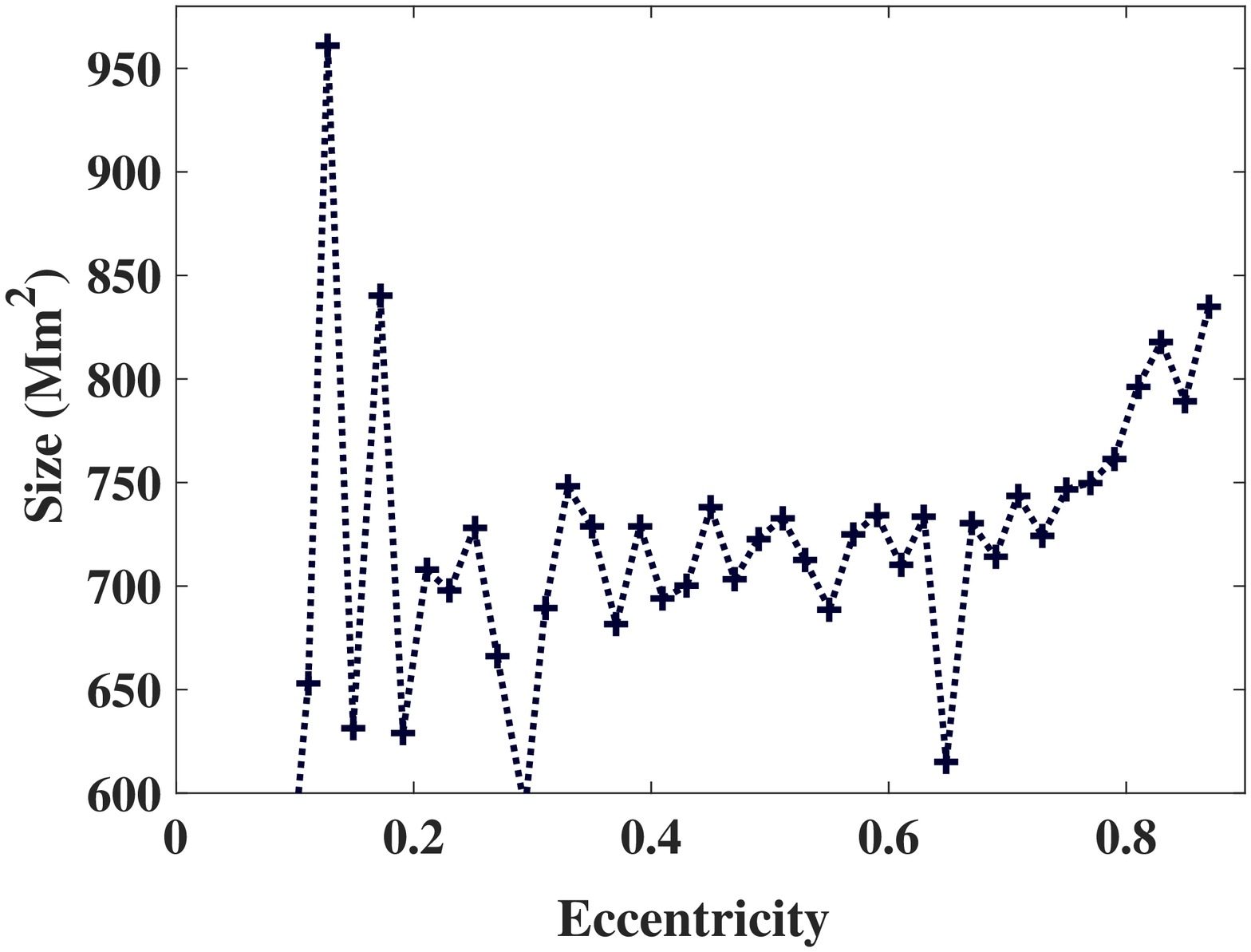}}
\centerline{\includegraphics[height=6cm]{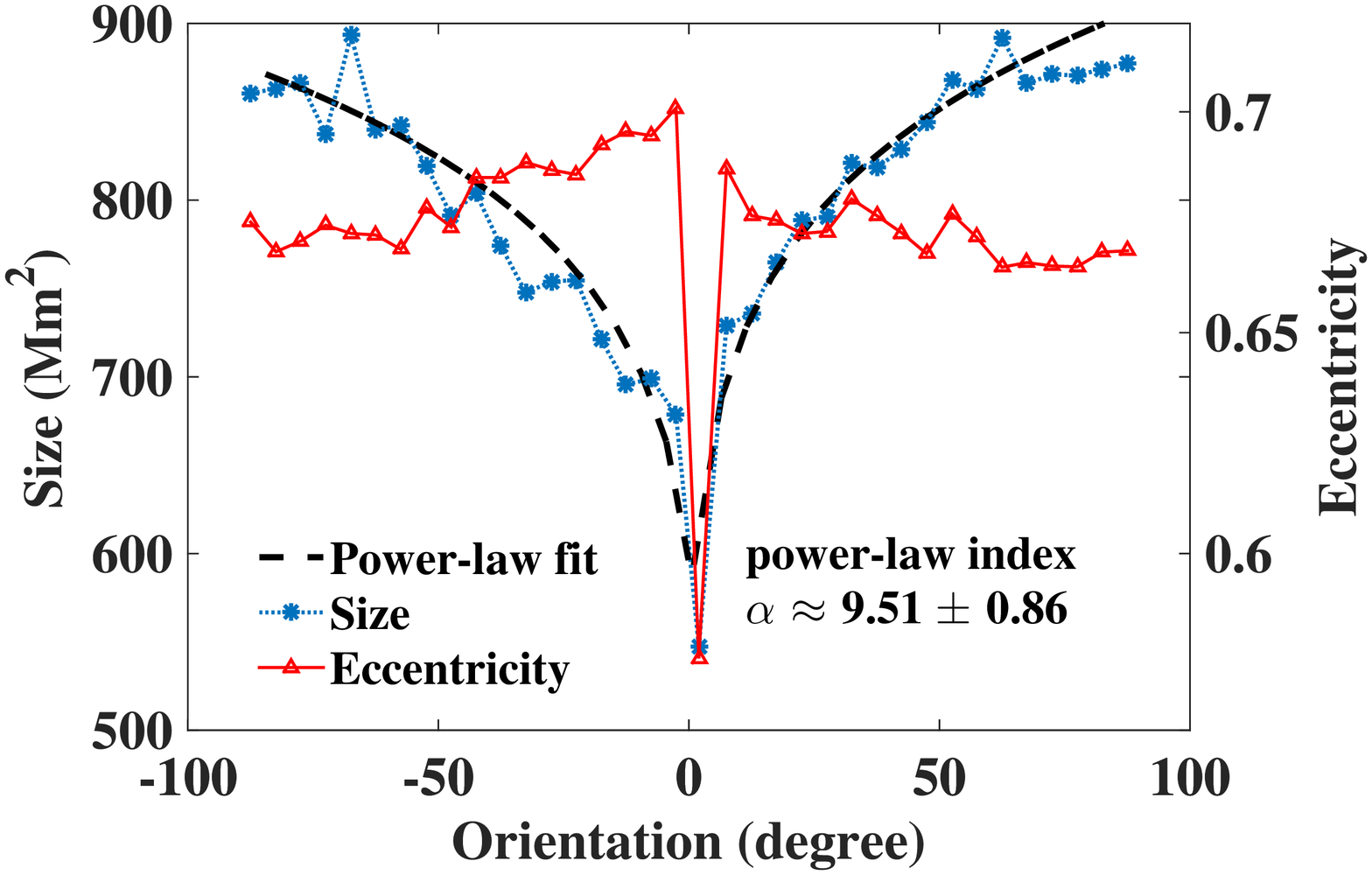}}
\caption{
Relationship between the size of cells (Mm$^2$) and their eccentricity (upper panel). The mean values for each bin (0 – 0.02, 0.02 – 0.04, etc.) for cells have been displayed. The relationship between the size of cells (Mm$^2$) and the orientation (degree) as mean values for each bin (0 – 5, 5 – 10, etc.) shows that the smaller sizes have no preference direction on the Sun (lower panel, blue line). The relationship between the eccentricity of cells and their orientation (degree) for mean values of each bin as mentioned are shown (lower panel, red line).}\label{fig7}
\end{figure}

\begin{figure}[ht!]
\centerline{\includegraphics[height=7.5cm]{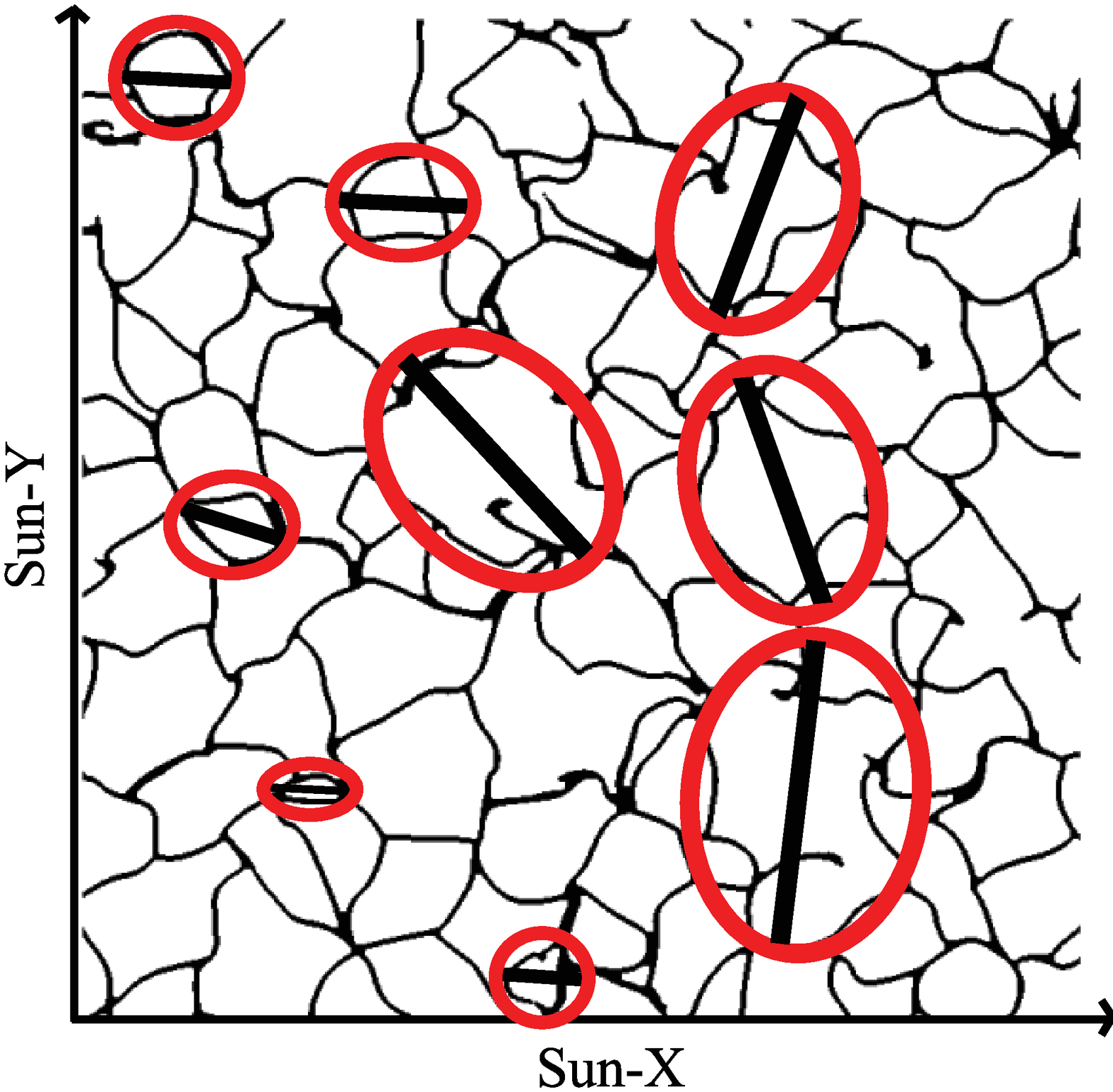}}
\caption{The fitted ellipses to the cells and their major axis. The orientations of the cells $-$ the angle between the major axis of the ellipse and the Sun-X (west-east direction) are ranged between -90$^{\circ}$ to +90$^{\circ}$.} \label{fig8}
\end{figure}

\section{Conclusions}\label{Con}
We used the ball-tracking method, edge sharpening technique, and box-counting algorithm to study the morphological parameters of photospheric supergranular cells. The code identified the number of 53651 individual cells from $900$ data cubes including SDO/HMI continuum images during the years 2011 to 2015. To avoid the projection effect (sphere to plan projection), $B_0$ angle evolution effect, variation of the solar radius during time, a box with the size of $350^{\prime\prime}\times350^{\prime\prime}$ around the central equatorial region (Fig. \ref{fig1}, red box) were studied. Since the fractal dimensions fluctuation is around 1.8 in each frame, it implies that the supergranulation pattern occur with self-similarities.

Size-frequency distribution of supergranules follows the log-normal function similar to that of obtained for granules \citep{Berrilli2002, Javaherian2014}. The eccentricity distributions of the cells possibly did not undergo the changes affected by high-activity or low-activity years. The power exponent of orientation distributions and the parameters of velocity-distributions don't vary significantly during the years 2011 to 2015. The orientation distribution of cells indicates the power-law behavior. Cells with smaller sizes ($<600~Mm^{2}$) have small angle of the orientations from $0^{\circ}$ to $5^{\circ}$ on the solar surface (Fig. \ref{fig7}, lower panel). The cells eccentricities fluctuate approximately around 0.66 (Fig. \ref{fig7}, upper panel). The power-law relation between size ($S$) and orientation ($O$) of the cells can be expressed by \textbf{$|O| \propto S^{9.5}$}. The results show that most of the small cells have small values of orientations (Fig. \ref{fig7}), and their average eccentricities are larger than 0.55 supporting non-roundish shapes. We concluded that the small cells align with the solar rotational velocity, while the larger ones are mostly orientated towards to the rotational axis. It seems that the solar rotation has not enough force to rotate the large cells along its rotational velocity. More quantitative studies are needed to investigate the influence of both differential rotation and photospheric magnetic field on cells orientations.

\newpage
\section*{Appendix A: Angle measurements and error analysis}
For a binary image $f(x,y)$, the area of an object $A = \Sigma_x \Sigma_y f(x,y)$ is a summation of pixels labeled one. So, we can introduce the centroid of object by following equations
\begin{equation}\label{centroid}
x_{CI} =\frac{1}{A} \Sigma_x \Sigma_y xf(x,y),~~~y_{CI} =\frac{1}{A} \Sigma_x \Sigma_y yf(x,y),~~~~~~~~~~~~~~~~~~~~~~~~~~~~~~~~~~~~~~~~~~~~~~~~~~~~~~~~~~~\nonumber
\end{equation}
where $x_{CI}$ and $y_{CI}$ are the coordinates of the intensity-weighted centroid. For an image, the central moments $\mu_{pq}$ is expressed as \citep{EmreCelebi2005}
\begin{equation}\label{moment}
\mu_{pq} = \Sigma_x \Sigma_y (x - x_{CI})^p (y - y_{CI})^q f(x,y).~~~~~~~~~~~~~~~~~~~~~~~~~~~~~~~~~~~~~~~~~~~~~~~~~~~~~~~~~~~~~~~~~~~~~~~~~~~~~\nonumber
\end{equation}
The angle $\phi$ between the major axis of an object and horizontal axis (positive $x$-axis) is given as \cite[e.g.,][]{Stoj}
\begin{equation}\label{angle}
\phi  = \frac{1}{2} \arctan\left( \frac{2\mu_{11}}{\mu_{20} - \mu_{02}} \right).~~~~~~~~~~~~~~~~~~~~~~~~~~~~~~~~~~~~~~~~~~~~~~~~~~~~~~~~~~~~~~~~~~~~~~~~~~~~~~~~~~~~~~
~~~~~~~~~~~~~~~~~~~~~~~~~~~~~ \nonumber
\end{equation}
The error of the angle $\phi$ as a function of the moments can be obtained by the \emph{error propagation} method \citep[e.g.,][]{Hughes,Mumford} defined as follows
\begin{equation}\label{errormu}
\Delta\mu_{pq} (x,y) = \pm \sqrt{\left(\Delta x \frac{\partial \mu_{pq}}{\partial x}\right)^2 + \left(\Delta y \frac{\partial \mu_{pq}}{\partial y}\right)^2},~~~~~~~~~~~~~~~~~~~~~~~~~~~~~~~~~~~~~~~~~~~~~~~~~~~~~~~~~~~~~~~~~~~~~~~~~~~~~~~~~~~~~\nonumber
\end{equation}
where $\Delta x$ and $\Delta y$ are spatial resolution in $x$ and $y$ direction, respectively. Since in our analysis, $\Delta x = \Delta y$, and are equal to one (pixel), the error of moments of interest takes the following form
\begin{eqnarray}\label{errorallmu}
\Delta\mu_{11} (x,y) = \pm ~~~~~~~~~~~~~~~~~~~~~~~~~~~~~~~~~~~~~~~~~~~~~~~~~~~~~~~~~~~~~~~~~~~~~~~~~~~~~~\nonumber \\
\sqrt{\left( \Sigma_Y Y^2 f^2(x,y) \right) + \left( \Sigma_X X^2 f^2(x,y) \right)}, ~~~~~~~~~~~~~~~~~~~~~~~~~~~~~~~~~~~~~~~~\nonumber \\
\Delta\mu_{20} (x,y) = \pm \sqrt{ \Sigma_X 4 X^2 f^2(x,y)}, ~~~~~~~~~~~~~~~~~~~~~~~~~~~~~~~~~~~~~~~~~~~~~~~~~~~~~\nonumber \\
\Delta\mu_{02} (x,y) = \pm \sqrt{ \Sigma_Y 4 Y^2 f^2(x,y)}, ~~~~~~~~~~~~~~~~~~~~~~~~~~~~~~~~~~~~~~~~~~~\nonumber
\end{eqnarray}
where $X = (x - x_{CI})$ and $Y = (y - y_{CI})$. So, the error of angle is obtained as follows
\begin{eqnarray}\label{errordeltathta}
\Delta \phi (\mu_{11},\mu_{20},\mu_{02}) = \pm ~~~~~~~~~~~~~~~~~~~~~~~~~~~~~~~~~~~~~~~~~~~~~~~~~~~~~~~~~~~~~~~~~~~~~~~~~~~~~~~~~~~~~~~~~~~~~~~~~~~~~~~~~~~~~~~~\nonumber \\
\sqrt{\left( \Delta \mu_{11} \frac{\partial \phi}{\partial \mu_{11} } \right)^2 + \left( \Delta \mu_{20} \frac{\partial \phi}{\partial \mu_{20} } \right)^2 + \left( \Delta \mu_{02} \frac{\partial \phi}{\partial \mu_{02} } \right)^2},~~~~~~~~~~~~~~~~~~~~~~~~~~~~~~~~~~~~~~~~~~~~~~~~~~~~~~~~~~~~~~~~~~~~~~~~~~~~~\nonumber
\end{eqnarray}
where the partial derivative are expanded as
\begin{eqnarray}\label{errortheta}
\frac{\partial \phi}{\partial \mu_{11}} = \frac{\mu_{20} - \mu_{02}}{(\mu_{20} - \mu_{02})^2 + 4\mu_{11}^2},~~~~~~~~~~~~~~~~~~~~~~~~~~~~~~~~~~~~~~~~~~~~~~~~~~~~~~~~~~~~~~~~~~~~~~~~~~~~~~~~~~~~~~~~~~~~~~~~~~~~~~~ \nonumber \\
\frac{\partial \phi}{\partial \mu_{20}} = \frac{- \mu_{11}}{(\mu_{20} - \mu_{02})^2 + 4\mu_{11}^2},~~~~~~~~~~~~~~~~~~~~~~~~~~~~~~~~~~~~~~~~~~~~~~~~~~~~~~~~~~~~~~~~~~~~~~~~~~~~~~~~~~~~~~~~~~~~~~~~~~~~~~~ \nonumber \\
\frac{\partial \phi}{\partial \mu_{02}} = \frac{ \mu_{11}}{(\mu_{20} - \mu_{02})^2 + 4\mu_{11}^2}.~~~~~~~~~~~~~~~~~~~~~~~~~~~~~~~~~~~~~~~~~~~~~~~~~~~~~~~~~~~~~~~~~~~~~~~~~~~~~~~~~~~~~~~~~~~~~\nonumber
\end{eqnarray}
As an example, we created 2-D binary form of artificial image \citep[mimicking data:][]{Javaherian2014} to test the validity of computing moments and estimate the error of orientation angle (Fig. \ref{fig9}). Using above-mentioned equations, the angle $\phi$ and the error value are obtained to be $31.6231\pm0.0032$ degrees.

\begin{figure}[ht!]
\centerline{\includegraphics[height=4.5cm]{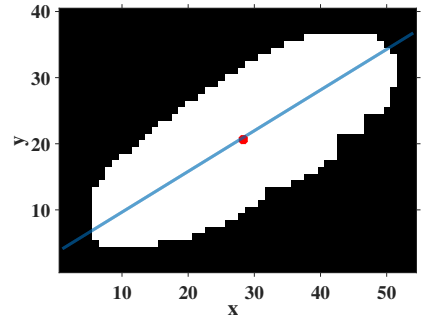}}
\caption{An artificial binary image with $54\times40$ pixels including an object described by a function $f(x,y)$. The red point is representative of the intensity-weighted centroid. The angle between the major axis of the object and the east-west direction is $31.6231\pm0.0032$ degrees.} \label{fig9}
\end{figure}




\newpage

\begin{thebibliography}{}
\expandafter\ifx\csname natexlab\endcsname\relax\def\natexlab#1{#1}\fi

\bibitem[{Alipour \& Safari(2015)}]{Alipour2015}
Alipour, N., \& Safari, H. 2015, ApJ, 807, 175

\bibitem[{Arish {et~al.}(2016)}]{Arish2016}
Arish, S., Javaherian, S., Safari, H. \& Amiri, A. 2016, Solar Phys, 291, 1209

\bibitem[{Aschwanden(2010)}]{Aschwanden2010}
Aschwanden, M.~J. 2010, Sol. Phys., 262, 235

\bibitem[{Aschwanden(2011)}]{Aschwanden2011}
Aschwanden, M.~J. 2011, Self-Organized in Astrophysics, Springer Heidelberg Dordrecht, London, UK, 257

\bibitem[{Aschwanden(2015)}]{Aschwanden2015}
Aschwanden, M.~J. 2015, ApJ, 814, 19


\bibitem[{Bazargan {et~al.}(2008)}]{Bazar}
Bazarghan, M., Safari, H., Innes, D.~E., Karami, E., \& Solanki, S.~K. 2008, A\&A, 492, L13

\bibitem[{Berrilli {et~al.}(2002)}]{Berrilli2002}
Berrilli, F., Consolini, G., Pietropaolo, E., Caccin, B., Penza, V., \& Lepreti, F. 2002, A\&A, 381, 253

\bibitem[{Boomgaard \& Balen(1992)}]{Boomgaard1992}
Boomgaard, R.~V.~D., \& Balen, R.~V. 1992, Computer Vision, Graphics, and Image Processing: Graphical Models and Image Processing,
  54(3), 254

\bibitem[{Del Moro {et~al.}(2007)}]{Del Moro2007}
Del Moro, D., Giordano, S., \& Berrilli, F. 2007, A\&A, 472, 599

\bibitem[{Emre Celebi \& Alp Aslandogan (2005)}]{EmreCelebi2005}
 Emre Celebi, M., \& Alp Aslandogan, Y. 2005, Conference: Coding and Computing (ITCC'05) International Conference on Information Technology. December, Las Vegas, NV, USA.

\bibitem[{Everitt(2002)Everitt}]{Everitt2002}
Everitt, B.~S. 2002, The Cambridge Dictionary of Statistics, Second edition, Cambridge University Press, New York, USA, 53, 304

\bibitem[{Farhang {et al.}(2018)}]{SafariFarhad}
Farhang, N., Safari, H., \& Wheatland, M. 2018, accepted for publication in ApJ

\bibitem[{Fisher \& Welsch(2008)}]{Fisher2008}
Fisher, G.~H., \& Welsch, B.~T. 2008, ASP Conference Series, 383, 373

\bibitem[{Gonzalez {et~al.}(2008)}]{Gonzalez2009}
Gonzalez, R.~C., Woods, R.~E. 2008, Digital Image Processing, Pearson Prentice Hall,
Upper Saddle River, New Jersey 07458, 631 - 633, 666

\bibitem[{Honarbakhsh {et~al.}(2016)}]{Honarbakhsh}
Honarbakhsh, L., Alipour, N., \& Safari, H. 2016, Sol. Phys., 291, 941

\bibitem[{Hughes \& Hase(2010)}]{Hughes}
Hughes, I.~G. \& Hase, T. P. A. 2010, Measuremants and Their Uncertainities,
First edition, Oxford University Press, New York, 37-44

\bibitem[{Javaherian {et~al.}(2014)}]{Javaherian2014}
Javaherian, M., Safari, H., Amiri, A., \& Ziaei, S. 2014, Sol. Phys., 289, 3969

\bibitem[{Javaherian {et~al.}(2017)}]{Javaherian2017}
Javaherian, M., Safari, H., Dadashi, N., \& Aschwanden, M.~J. 2017, Sol. Phys., 292, 164

\bibitem[{Innes {et~al.}(2009)}]{Innes}
Innes, D.~E., Genetelli, A., Attie, R., \& Potts, H. E. 2009, Sol. Phys., 495, 3191

\bibitem[{Meunier {et~al.}(2007)}]{Meunier2007}
Meunier, N., Roudier, T., \& Tkaczuk, R. 2007c, A\&A, 466, 1123

\bibitem[{Meunier {et~al.}(2008)}]{Meunier2008}
Meunier, M., Roudier, T., \& Rieutord, M. 2008, A\&A, 488, 1109

\bibitem[{Molteno(1993)}]{Molteno1993}
Molteno, T.~C.~A. 1993, Physical Review E, 48, R3263

\bibitem[{Mumford(2017)}]{Mumford}
Mumford, J.~R. 2017, Error Propagation Reference, John Hopkins University, Physics and Astronomy, Basic Physics Laboratory, 2-4

\bibitem[{Newman(2006)}]{Newman2006}
Newman, M.~E.~J. 2006, Contemporary Physics E, 46, 323

\bibitem[{Papadimitriou {et~al.}(2006)}]{Papadimitriou2006}
Papadimitriou, S., Sun, J., \& Philip, S. 2006, Conference: Proceedings of the 6th IEEE International Conference on Data Mining. December, Hong Kong, China

\bibitem[{Potts {et~al.}(2004)}]{Potts2004}
Potts, H.~E., Barrett, R., \& Diver, D.~A. 2004, A\&A, 424, 253

\bibitem[{Potts \& Diver(2008)}]{Potts2008}
Potts, H.~E., \& Diver, D.~A. 2008, Sol. Phys., 248, 263

\bibitem[{Press {et~al.}(2007)}]{Press2007}
Press, W.~H., Teukolsky, S.~A., Vetterling, W.~T., \& Flannery, B.~P. 2007, Numerical Recieps: The Art of Scientific Computing,
Third edition, Cambridge university Press, New York, USA, 724

\bibitem[{Priest(2014)Priest}]{Priest2014}
Priest, E., 2014, Magnetohydrodynamics of the Sun, First edition, Cambridge University of the Sun, New York, USA, 21, 22

\bibitem[{Rincon {et~al.}(2017)}]{Rincon2017}
Rincon, F., Roudier, T., Schekochihin, A.~A., \& Rieutord, M. 2017, A\&A, 599, A69

\bibitem[{Roudier {et~al.}(2013)}]{Roudier2013}
Roudier, T., Rieutord, M., Prat, V., {et~al.} 2013, A\&A, 552, A113

\bibitem[{Roudier {et~al.}(2014)}]{Roudier2014}
Roudier, T., Svanda, M., Rieutord, M., {et~al.} 2014, A\&A, 567, A138


\bibitem[{Ryutova {et~al.}(2003)}]{Ryutova}
Ryutova, M., Tarbell, T.~D., Shine, R. 2003, Sol. Phys., 213, 231

\bibitem[{Ryutova (2015)}]{Margarita}
Ryutova, M. 2015, Physics of Magnetic Flux Tubes, Springer-Verlag GmbH Berlin Heidelberg, 12

\bibitem[{Scherrer {et~al.}(1995)}]{Scherrer}
Scherrer, P.~H., Bogart, R.~S., Bush, R.~I., {et~al.} 1995, Sol. Phys., 162, 129

\bibitem[{Schou {et~al.}(2012)}]{Schou2012}
Schou, J., Scherrer, P.~H., Bush, R.~T., {et~al.} 2012, Sol. Phys., 275, 229

\bibitem[{Simon \& Weiss(1968)}]{Simon1968}
Simon, G.~W., \& Weiss, N.~O. 1968, Zeitschrift für Astrophysik, 69, 435

\bibitem[{Stangalini(2013)Stangalini}]{Stangalini2013}
Stangalini, M. 2013, A\&A, 561, L6

\bibitem[{Stojmenovic \& Nayak (2007)}]{Stoj}
Stojmenovic, M., Nayak, A., 2007, Direct Ellipse Fitting and Measuring Based on Shape Boundaries, Springer,
Berlin, Heidelberg, 221-223

\bibitem[{Tajfirouze \& Safari (2012)}]{Taj}
Tajfirouze, E., \& Safari, H. 2012, ApJ, 744, 113

\bibitem[{Tian {et~al.}(2010)}]{Tian2010}
Tian, H., Potts, H.~E., Marsch, E., Attie, R., \& He, J. 2010, A\&A, 519, 10

\bibitem[{Tian {et~al.}(1989)}]{Title1989}
Title, A.~M., Tarbell, T.~D., Topka, K.~P., {et~al.} 1989, ApJ, 336, 475

\bibitem[{Wachter {et~al.}(2012)}]{Wachter2012}
Wachter, R., Schou, J., Rabello-Soares, M.~C., Miles, J.~W., Duvall Jr. T.~L., \& Bush, R.~I. 2012, Sol. Phys., 275, 261

\bibitem[{Yousefzadeh {et~al.}(2016)}]{Yousefzadeh2016}
Yousefzadeh, M., Safari, H., Attie, R., \& Alipour, N. 2016, Sol. Phys., 291, 29
\end{thebibliography}
\end{document}